\documentclass[preprintnumbers,amsmath,amssymb,floatfix,10pt,prd,twocolumn,superscriptaddress,nofootinbib]{revtex4-2}
\usepackage{latexsym}
\usepackage{epsfig}
\usepackage{epstopdf}
\usepackage{graphicx}
\usepackage{amssymb}
\usepackage{amsmath}
\usepackage{amsfonts}
\usepackage{subfigure}
\usepackage{dcolumn}
\usepackage{bm}
\usepackage{color}
\usepackage{comment}
\usepackage[utf8]{inputenc}
\usepackage[dvipsnames]{xcolor}
\usepackage{hyperref}
\hypersetup{colorlinks=true,linkcolor=blue,citecolor=red,urlcolor=magenta}
\usepackage{orcidlink}
\usepackage{booktabs}

\begin{document}
\title{\bf{Testing the Universality of RAR Model with Constraints on Fermionic Dark Matter for M31}}
	
\author{Aqsa Yasmin \orcidlink{0009-0007-7368-8685}}
\email{aqsayasmin875@gmail.com}
\affiliation{Department of Physics and Astronomy, School of Natural Sciences, National University of Sciences and Technology, Sector H-12, Islamabad, Pakistan}
	
\author{Ibrar Hussain}
\email{ibrar.hussain@seecs.nust.edu.pk}
\affiliation{School of Electrical Engineering and Computer Science, National University of Sciences and Technology, Sector H-12, Islamabad, Pakistan}	
	
\author{Mubasher Jamil}
\email{mjamil@sns.nust.edu.pk}
\affiliation{Department of Mathematics, School of Natural Sciences, National University of Sciences and Technology, Sector H-12, Islamabad, Pakistan}

\begin{abstract}The Ruffini--Argüelles--Rueda (RAR) model describes dark matter as a self-gravitating system of massive fermions in thermodynamic equilibrium within general relativity. It suggests a dual morphology in galaxies: a highly degenerate quantum core mimicking a central supermassive black hole, and a classical halo that explains galactic rotation curves. Earlier applications of the model to the Milky Way restricted the fermion mass to $mc^{2}=48\text{--}345~\text{keV}$, and the orbits of the S-stars around Sagittarius~A$^{\ast}$ (Sgr~A$^{\ast}$) favored $mc^{2}=56~\text{keV}$, whereas a recent thermodynamic stability analysis allows only $mc^{2}\approx194\text{--}387~\text{keV}$. In this work, we apply the RAR model to the Andromeda galaxy (M31), which hosts a dark compact object of approximately $1.4\times10^{8}~M_{\odot}$ within $0.19~\text{pc}$, roughly 33 times the mass of Sgr~A$^{\ast}$. We solve the relativistic equilibrium equations for $mc^{2}=6\text{--}60~\text{keV}$. For all these masses, the fermionic core lies within $0.19~\text{pc}$ and provides the central mass without a black hole. With a bulge and disk added, the total rotation curve is statistically acceptable ($\chi^{2}/N_{\text{dof}}\approx1.18$) but lies below the disk data at $3\text{--}11~\text{kpc}$, an offset attributed to the baryonic models we adopted. The Navarro--Frenk--White (NFW) and Burkert profiles fitted to the same data give a lower $\chi^{2}$ but require an additional black hole. The stability of the M31 core against gravitational collapse limits the fermion mass to $mc^{2}\lesssim60~\text{keV}$. With the earlier Milky Way range, a single fermion can therefore describe both galaxies only for $mc^{2}=48\text{--}60~\text{keV}$, which includes $56~\text{keV}$. With the thermodynamic range, no fermionic core can supply the central mass of M31, so M31 would have to host a black hole.
	
{\textbf{keywords}}: Dark Matter; Fermionic Dark Matter; Galaxies; Andromeda M31 Galaxy; Supermassive Black Holes; Galactic Halos; Fermions; Rotation Curves.
\end{abstract}
\maketitle
\date{\today}
	
\section{Introduction}\label{S1}
General relativity predicted black holes as the extreme spacetime geometries where gravitational fields prevent even the escape of light \cite{Einstein:1915ca,Schwarzschild:1916uq}. For nearly a century, their existence was only tested indirectly, such as through the dynamics of accretion disks and stellar orbits, until gravitational waves from black holes merger \cite{LIGOScientific:2016aoc} were detected in 2015. Two questions arise while discussing a black hole: what does it look like, and are the compact objects at the centers of galaxies black holes? If not, what are they made of? To answer the first question, Synge \cite{1966MNRAS.131..463S} and Bardeen \cite{Bardeen:1973tla} developed mathematical frameworks to determine the image of a black hole in terms of its shape and size, known as its shadow. Eventually, the predictions were put to test when the Event Horizon Telescope (EHT) collaboration presented the first images of the supermassive black hole M87$^{\ast}$ in 2019 \cite{EventHorizonTelescope:2019dse,EventHorizonTelescope:2019uob,EventHorizonTelescope:2019jan,EventHorizonTelescope:2019ths,EventHorizonTelescope:2019pgp,EventHorizonTelescope:2019ggy}, followed by Sagittarius~A$^{\ast}$ (Sgr~A$^{\ast}$) at the center of the Milky Way in 2022 \cite{EventHorizonTelescope:2022wkp,EventHorizonTelescope:2022apq,EventHorizonTelescope:2022wok,EventHorizonTelescope:2022exc,EventHorizonTelescope:2022urf,EventHorizonTelescope:2022xqj}, showing a dark central region surrounded by a bright ring of emission from hot plasma. The images are consistent with the predictions for black holes and test general relativity in the strong-field regime.

To answer the second question about the nature of these compact objects, alternative objects that could mimic black holes have been proposed, such as boson and Proca stars fundamentally made up of bosons \cite{Kaup:1968zz,Ruffini:1969qy,Brito:2015pxa}, fuzzballs made of one-dimensional vibrating strings and D-branes \cite{Mathur:2005zp}, and dense fermionic dark matter cores \cite{2014IJMPD..2342020A,2018PDU....21...82A}. The images generated by simulations of such fermionic cores show a central brightness depression which is surrounded by a ring-like structure, but no photon ring, distinguishing them from black holes \cite{Pelle:2024eyt}. Jusufi et al. \cite{PhysRevD.100.044012} studied the shadow of a black hole surrounded by the dark matter halo of M87, and Nampalliwar et al. \cite{Nampalliwar:2021tyz} investigated the properties of Sgr~A$^{\ast}$ immersed in dark matter.

The nature of dark matter \cite{2005PhR...405..279B} is one of the major unresolved problems and a mystery of modern physics. Cosmological observations indicate that only about 4.9\% of the total energy density of the Universe is ordinary matter, 26.4\% is dark matter, and 68.5\% is the so-called dark energy \cite{Planck:2018vyg}. The $\Lambda$ cold dark matter ($\Lambda$CDM) model has been highly successful in describing the large-scale structure formation, but it faces challenges on galactic scales, in particular the core-cusp problem and the missing satellites problem \cite{2017ARA&A..55..343B}, motivating astronomers to work on alternative models which may solve these discrepancies. Another problem with the early formation and growth of black holes is that stellar-mass black holes form by gravitational core collapse of massive stars \cite{2003ApJ...591..288H} and provide seeds of $\sim10\text{--}100~M_{\odot}$ \cite{KAGRA:2021vkt}, whereas supermassive black holes have masses of $\sim10^{6}~M_{\odot}\text{--}10^{10}~M_{\odot}$ \cite{Kormendy:2013dxa}. The quasar ULAS J1342+0928 at redshift $z=7.54$ hosted a black hole of about $8\times10^{8}~M_{\odot}$ when the Universe was only $\sim690$ Myr old \cite{2018Natur.553..473B}, and the X-ray quasar UHZ1 at $z\approx10$ hosted a black hole of about $10^{7}\text{--}10^{8}~M_{\odot}$ only $\sim500$ Myr after the Big Bang \cite{Bogdan:2023ilu}. For stellar-mass seeds to grow to such masses would require nearly uninterrupted Eddington-limited accretion over the entire age of the Universe at that epoch \cite{2020ARA&A..58...27I}.

Ruffini, Argüelles, and Rueda \cite{2015MNRAS.451..622R} proposed a model, known as the RAR model, that addresses and proposes solutions for the problems mentioned above. The model treats dark matter as a self-gravitating system of massive fermions in thermodynamic equilibrium, incorporating general relativistic effects, which are essential for cores approaching gravitational collapse. A consequence of the model is that it develops a core-halo morphology as a result of Pauli's exclusion principle: a dense, highly degenerate quantum core at the galactic center and a dilute halo that obeys classical Boltzmann statistics \cite{Ruffini:1969qy,1998MNRAS.296..569C,1988NCimB.101..369I}. Applications of the RAR model to other galaxies and to self-interacting dark matter have also been studied \cite{2016JCAP...04..038A,Arguelles:2015zef,Mavromatos:2016vbj,Krut:2023drx}. Argüelles et al. \cite{2018PDU....21...82A} applied the model to the Milky Way, adding the energy cutoff to the distribution function to account for the finite size of the galaxy, constraining the fermion mass to $mc^{2}=48\text{--}345~\text{keV}$. The lower bound requires the core to fit inside the S2 pericenter at $\sim6\times10^{-4}~\text{pc}$ \cite{2009ApJ...692.1075G}, and the upper bound requires the fermionic core of $\sim4\times10^{6}~M_{\odot}$ to stay below the critical mass for gravitational collapse. They showed that such cores can provide an alternative to the black hole interpretation of Sgr~A$^{\ast}$. Moreover, if a quantum core accretes enough baryonic matter and exceeds the critical mass limit, it collapses into a black hole \cite{Arguelles:2023kqw}, providing a channel for supermassive black hole formation that bypasses the mechanism of slow growth of the stellar-mass seeds \cite{2023MNRAS.523.2209A}.

The stars orbiting Sgr~A$^{\ast}$ were studied and this narrowed down the fermion mass within the given range. For $mc^{2}=56~\text{keV}$, the core fits the orbits of S2 and G2, fitting G2 better than a central black hole \cite{2020A&A...641A..34B}. The value $mc^{2}=56~\text{keV}$ is said to be the best-fit fermion mass for the Milky Way because its solution fits the orbits of the 17 best-resolved S-stars and the Milky Way rotation curves \cite{Becerra-Vergara:2021gmx}. From the Bayesian analysis of the S2 orbit, $mc^{2}=56~\text{keV}$ is preferred only slightly over $mc^{2}=300~\text{keV}$ \cite{Crespi:2025ygl}. In addition, the thermodynamic stability of these core-halo solutions was studied by Krut et al. \cite{Krut:2025jju} and found stable solutions for the Milky Way only for $mc^{2}\approx194\text{--}387~\text{keV}$, the lower bound came from the thermodynamic stability of a core of $4.2\times10^{6}~M_{\odot}$, and the upper bound from the Oppenheimer--Volkoff limit. For $mc^{2}=56~\text{keV}$, they found no stable solution to account for Sgr~A$^{\ast}$. 

If the RAR fermions are the dark matter particles, a single fermion mass must describe every galaxy, while the free parameters of the model (the temperature, degeneracy, and cutoff parameters) vary from one galaxy to another. The Andromeda galaxy (M31) is the nearest massive spiral galaxy, providing a critical testing ground for this framework. It has a central mass of $\sim1.4\times10^{8}~M_{\odot}$ within $0.19~\text{pc}$ \cite{2005ApJ...631..280B}, roughly 33 times the mass of Sgr~A$^{\ast}$, and a rotation curve extending to $\sim120~\text{kpc}$ \cite{2024MNRAS.528.2653Z}. Because the critical mass of a degenerate fermionic core varies as the inverse square of the fermion mass, $M_{\text{cr}}\propto m^{-2}$ \cite{2014IJMPD..2342020A}, a massive core reaches the critical mass at a lower fermion mass, which lowers the upper bound on the particle mass. The upper bounds for the Milky Way and M31 are $mc^{2}\approx 345~\text{keV}$ \cite{2018PDU....21...82A} and $mc^{2}\approx 60~\text{keV}$ (this work), respectively. Both bounds assume that the central dark object is a fermionic core rather than a black hole. M31's innermost mass constraint lies at $0.19~\text{pc}$, more than 300 times farther out than the Milky Way's S2 pericenter, so the lower bound on the fermion mass for M31 cannot be improved, we therefore consider masses down to $mc^{2}\approx 6~\text{keV}$. To our knowledge, the RAR model has not been applied to M31 as an individual system \cite{2023Univ....9..197A}.

In this work, we apply the RAR model to M31 for fermion masses in the range $mc^{2}=6\text{--}60~\text{keV}$ to test the universality of the dark matter fermions. We fix the free parameters with the central mass and two halo masses from the rotation curve. We then obtain the dark matter density profiles and rotation curves, add the bulge and disk, and test the total rotation curve against the observational data with a $\chi^{2}$ analysis. We also compare the results with the Navarro--Frenk--White (NFW) and Burkert profiles. We find the core stability of M31 against gravitational collapse limits the fermion mass to $mc^{2}\lesssim60~\text{keV}$. For the fermion masses we consider, the core provides the central mass of M31 without a black hole, and the total rotation curve is statistically acceptable. Comparing the solutions for $mc^{2}=56~\text{keV}$, we see that the M31 core is roughly 34 times more massive than the Milky Way core but 4 times smaller, hence much closer to gravitational collapse. With the earlier Milky Way range, a single fermion can describe both the Milky Way and M31 only for $mc^{2}=48\text{--}60~\text{keV}$, which contains the Milky Way best-fit value $mc^{2}=56~\text{keV}$. For the thermodynamic range of Krut et al. \cite{Krut:2025jju}, the critical core mass is below the central mass of M31, so the dark compact object must be a black hole. The upper bound of the mass range scales with the central mass, so a more precise measurement of the central mass of M31 can test whether a single fermion describes both galaxies.

The paper is organized as follows. In section \ref{S2}, we summarize the theoretical framework of the RAR model. Section \ref{S3} gives the application of the RAR model to M31: we set the boundary conditions, obtain the dark matter density profiles and rotation curves for different fermion masses, compute the total rotation curve with baryons and dark matter, compare the solution for $mc^{2}=56~\text{keV}$ with that of the Milky Way, and determine what the two Milky Way mass ranges imply for M31. The focus is to see how the dark matter profiles vary with the particle mass. In section \ref{S4}, we present the statistical kinematic analysis, and in section \ref{S5}, we compare the RAR model with the NFW and Burkert profiles. Section \ref{S6} summarizes our results and concludes with future prospects.

\section{The Ruffini--Argüelles--Rueda Model}\label{S2}
The RAR model \cite{2015MNRAS.451..622R} describes the dark matter distribution in galaxies, with a central compact quantum core that could be an alternative to a black hole, and a classical halo. It considers a self-gravitating system of neutral, spin-1/2 fermions in thermodynamic equilibrium, obeying quantum statistics within general relativity. The density $\rho$ and pressure $P$ of the fermionic dark matter follow from Fermi--Dirac statistics \cite{2016JCAP...04..038A, 2018PDU....21...82A}. To account for the finite size of galaxies, an energy cutoff is introduced in the distribution function \cite{2018PDU....21...82A}, so that
\begin{equation}\label{eqn:rho}
\rho=\frac{2m}{h^{3}}\int^{\epsilon_{c}}_{0} f_{c}(p)\left[ 1+\frac{\epsilon(p)}{mc^{2}}\right]\ d^{3}p,
\end{equation}
and
\begin{equation}\label{eqn:P}
P=\frac{4}{3h^{3}}\int^{\epsilon_{c}}_{0} f_{c}(p)\left[ 1+\frac{\epsilon(p)}{mc^{2}}\right]^{-1}\left[ 1+\frac{\epsilon(p)}{2mc^{2}}\right]\epsilon\ d^{3}p,
\end{equation}
where $f_{c}$ is the distribution function with the energy cutoff, defined as
\begin{equation}\label{eqn:f}
f_{c}(\epsilon\leq\epsilon_{c})=\frac{1-e^{\frac{(\epsilon-\epsilon_{c})}{k_{B}T}}}{e^{\frac{(\epsilon-\mu)}{k_{B}T}}+1}, \qquad f_{c}(\epsilon>\epsilon_{c})=0,
\end{equation}
and the relativistic kinetic energy $\epsilon$ is defined as
\begin{equation}\label{eqn:ep}
\epsilon= \sqrt{c^{2}p^{2}+m^{2}c^{4}}-mc^{2}.
\end{equation}
In the above equations, $m$ is the fermion mass, $p$ is the momentum, $h$ is Planck's constant, $c$ is the speed of light, and $\mu$ is the chemical potential (with the rest energy of the particle subtracted off). Furthermore, $\epsilon_{c}$ is the cutoff energy, $k_{B}$ is the Boltzmann constant, and $T$ is the temperature \cite{2018PDU....21...82A}. The system has some parameters governing it: the temperature parameter $\beta$, the degeneracy parameter $\theta$, and the cutoff parameter $W$, defined as
\begin{equation}\label{eqn:fp}
\beta=\frac{k_{B}T}{mc^{2}}, \qquad \theta=\frac{\mu}{k_{B}T}, \qquad W=\frac{\epsilon_{c}}{k_{B}T}.
\end{equation}
The values of these parameters at the center, $\beta_{0}$, $\theta_{0}$, and $W_{0}$, are the three free parameters of the model for a given fermion mass. Anti-fermions are not considered in this model, which is valid for $k_{B}T\ll mc^{2}$ (i.e.\ $\beta\ll 1$). To describe the spacetime geometry, we adopt a static spherically symmetric metric
\begin{equation}\label{eqn:met}
ds^{2}=e^{\nu}c^{2}dt^{2}-e^{\lambda}dr^{2}-r^{2}d\vartheta^{2}-r^{2}\sin^{2} \vartheta d\phi^{2},
\end{equation}
where $\nu(r)$ and $\lambda(r)$ are the metric functions, with $e^{-\lambda(r)}=1-\frac{2GM(r)}{c^{2}r}$ and $M(r)$ is the dark matter mass enclosed within radius $r$. The Tolman \cite{PhysRev.35.904} and Klein \cite{RevModPhys.21.531} thermodynamic equilibrium conditions are considered
\begin{equation}\label{eqn:TK}
e^{\frac{\nu}{2}}T=\text{constant}, \qquad e^{\frac{\nu}{2}}(\mu+mc^{2})=\text{constant}.
\end{equation}
The cutoff condition is obtained from energy conservation along a geodesic
\begin{equation}\label{eqn:cc}
e^{\frac{\nu}{2}}(\epsilon+mc^{2})=\text{constant},\qquad (1+W\beta)=e^\frac{(\nu_{b}-\nu)}{2},
\end{equation}
where $\nu_{b}=\nu(r_{b})$ is the metric function at the boundary radius $r_{b}$ of the halo, at which $\epsilon_{c}(r_{b})=0$ and therefore $W(r_{b})=0$ \cite{2018PDU....21...82A}. The relativistic equilibrium equations of the model are
\begin{eqnarray}\label{eqn:eqnM}
\frac{d\hat{M}}{d\hat{r}}&=&4\pi \hat{r}^{2}\hat{\rho}, \\
\frac{d{\theta}}{d\hat{r}}&=&- \frac{1-\beta_{0}(\theta-\theta_{0})}{\beta_{0}}\frac{\hat{M}+4\pi \hat{P}\hat {r}^{3}}{\hat{r}^{2}(1-\frac{2\hat{M}}{\hat{r}})}, \\
\frac{d\nu}{d\hat{r}}&=&\frac{2(\hat{M}+4\pi \hat{P}\hat {r}^{3})}{\hat{r}^{2}(1-\frac{2\hat{M}}{\hat{r}})}, \\
\beta(\hat r)&=&\beta_{0}e^{\frac{\nu_{0}-\nu(\hat r)}{2}}, \\
W(\hat r)&=&W_{0}+\theta(\hat r)-\theta_{0}.
\end{eqnarray}
The equilibrium equations are made dimensionless with
\begin{equation}\label{eqn:eqnDim}
\hat{\rho}=\frac{G\ell^{2}\rho}{c^{2}}, \quad \hat{P}=\frac{G\ell^{2}P}{c^{4}}, \quad \hat{M}=\frac{GM}{c^{2}\ell}, \quad  \hat{r}=\frac{r}{\ell},
\end{equation}
where $\ell=2\pi^\frac{3}{2}\left(\frac{\hbar m_{p}}{m^{2}c}\right)$ and the Planck mass is given as $m_{p}=\sqrt{\frac{\hbar c}{G}}$. The circular velocity for the dark matter is given by \cite{2015MNRAS.451..622R}
\begin{equation}\label{eqn:DM}
v_{\text{DM}}(r)=\sqrt{\frac{GM(r)}{r-\frac{2GM(r)}{c^2}}}.
\end{equation}
The total rotation curve incorporates dark matter and baryons, so the total circular velocity is given as
\begin{equation}\label{eqn:bardark}
v^{2}_{\text{tot}}(r)=v^{2}_{\text{bulge}}(r)+v^{2}_{\text{disk}}(r)+v^{2}_{\text{DM}}(r),
\end{equation}
where $v_{\text{bulge}}$ and $v_{\text{disk}}$ are the circular velocities for bulge and disk regions. The initial and boundary conditions for the model, along with the bulge and disk models, are given in section \ref{S3}. Solving the system of equilibrium equations with the initial and boundary conditions, we obtain the core properties, density profiles, and rotation curves for different particle masses. These solutions provide a channel to check whether the quantum core is compact enough to undergo gravitational collapse into a black hole \cite{2014IJMPD..2342020A,2019PDU....24..278A}. In this work, the core mass is required to stay below the critical mass for gravitational collapse, $M_{\text{cr}}\propto m^{-2}$ \cite{2014IJMPD..2342020A}, there is an additional condition imposed by the thermodynamic stability of the core-halo solutions for the Milky Way \cite{Krut:2025jju} and we discuss its consequences for M31 in section \ref{S3}. The compact central cores offer a possible explanation for the early formation of supermassive black holes \cite{2023MNRAS.523.2209A}.

\section{Application to M31}\label{S3}
The nearest large spiral galaxy to the Milky Way \cite{2005MNRAS.356..979M}, M31 provides a test for the RAR model. We consider the fermion masses in the range $mc^{2}=6\text{--}60~\text{keV}$ and obtain the resulting solutions (core properties), density profiles, and rotation curves. 

\subsection{Boundary Conditions}
\begin{figure*}[t!]
	\centering
	\subfigure{\includegraphics[width=0.45\textwidth]{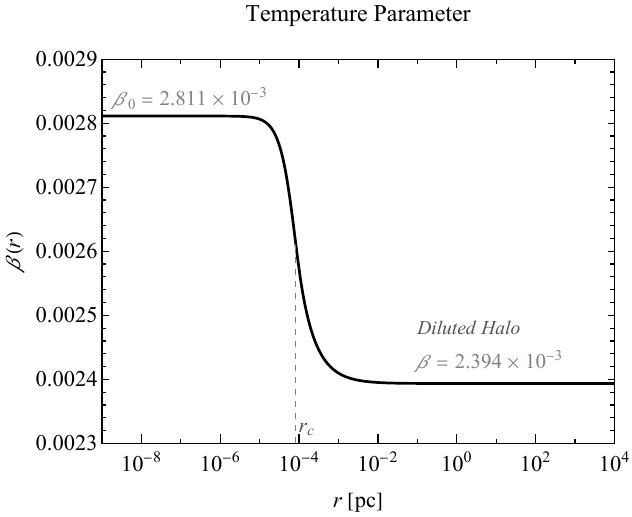}}~~~~
	\subfigure{\includegraphics[width=0.45\textwidth]{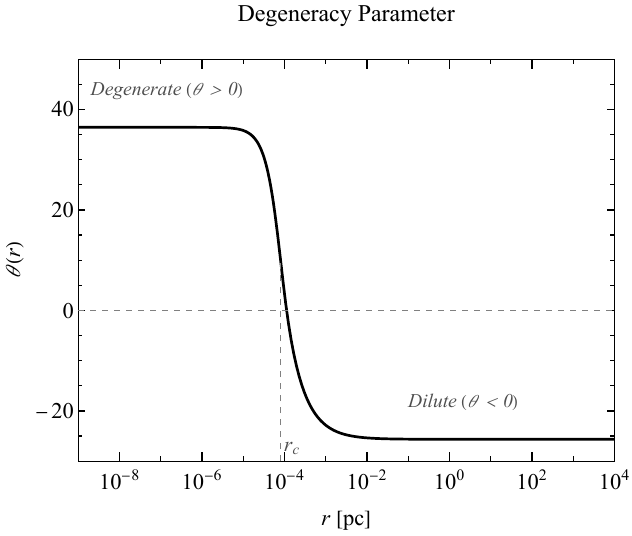}}\\
	\subfigure{\includegraphics[width=0.45\textwidth]{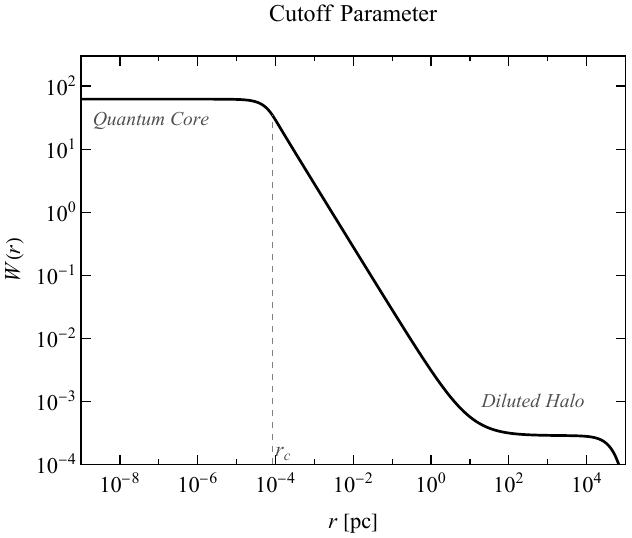}}
	\caption{The M31 RAR model plots for model parameters (solid black curves): \textbf{(a)} the temperature parameter $\beta$, \textbf{(b)} the degeneracy parameter $\theta$, \textbf{(c)} the cutoff parameter $W$ for fermion mass $60~\text{keV/c}^{2}$.}\label{fig:plot5}
\end{figure*}

The innermost radius within which the central mass of M31 is measured is $0.19~\text{pc}$ \cite{2005ApJ...631..280B}, no observations have improved this result yet \cite{2018ApJ...854..121L}. A fermion mass is excluded if its corresponding core radius exceeds this observational limit on the innermost radius, requiring $mc^{2}\gtrsim3.6~\text{keV}$. A fermion mass is also excluded if the central mass exceeds the critical mass for gravitational collapse, $M_{\text{cr}}\propto m^{-2}$, which requires $mc^{2}\lesssim60~\text{keV}$. Dynamical modeling of M31 gives the first boundary condition \cite{2005ApJ...631..280B}
\begin{equation}\label{eqn:bc1}
M_{\text{DM}}(0.19~\text{pc})=1.4^{+0.9}_{-0.3}\times10^{8}~M_{\odot}.
\end{equation} 
The remaining boundary conditions are derived from the rotation curve of Zhang et al. \cite{2024MNRAS.528.2653Z} by subtracting the baryonic contribution given by Geehan et al. \cite{2006MNRAS.366..996G}, we have
\begin{eqnarray}\label{eqn:bc2}
M_{\text{DM}}(31.23~\text{kpc})=(1.726\pm1.12)\times10^{11}~M_{\odot}, \\
M_{\text{DM}}(98.74~\text{kpc})=(7.338\pm3.73)\times10^{11}~M_{\odot}.
\end{eqnarray} 
The initial conditions at the center $(r=0)$ are $M(0)=0$, $\theta(0)=\theta_{0}$, $\nu(0)=0$, $\beta(0)=\beta_{0}$, and $W(0)=W_{0}$ \cite{2018PDU....21...82A}. Using these initial and boundary conditions, we solve the relativistic equilibrium system of equations. For the above boundary conditions, we have a set of free parameters for each fermion mass. The parameters are given in table \ref{table1}. In figure \ref{fig:plot5}, we show the profiles of the model parameters as functions of $r$ for $mc^{2}=60~\text{keV}$.
  
\subsection{Density Profiles}
\begin{figure*}[t!]
\centering
\subfigure{\includegraphics[width=0.55\textwidth]{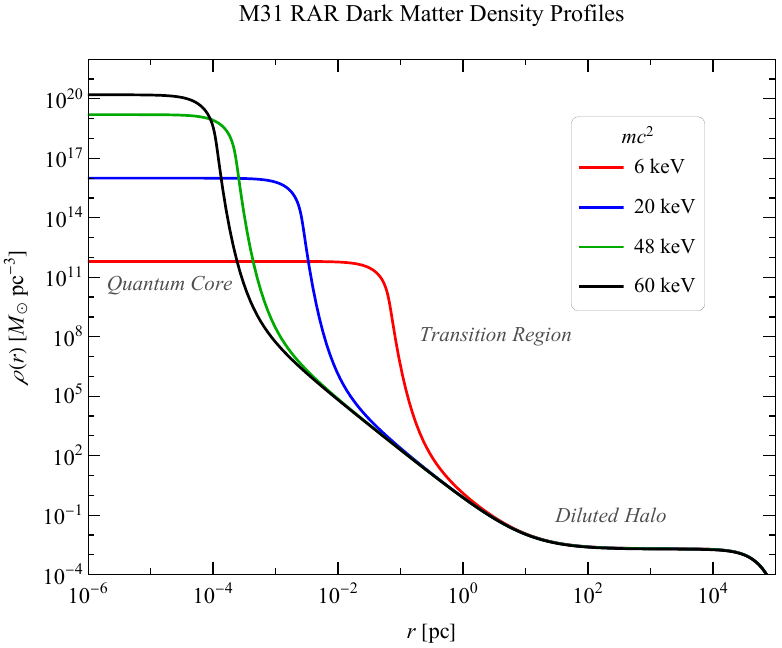}}
\caption{The M31 dark matter density profiles for fermion masses $6~\text{keV/c}^{2}$ (solid red curve), $20~\text{keV/c}^{2}$ (solid blue curve), $48~\text{keV/c}^{2}$ (solid green curve), and $60~\text{keV/c}^{2}$ (solid black curve) from $10^{-6}~\text{pc}$ to $10^{5}~\text{pc}$. The profiles show three regions: a quantum core, a transition region, and a classical halo.}\label{fig:plot1}
\end{figure*}

We solve the system of equilibrium equations and plot the density profiles $\rho(r)$ for $mc^{2}=6$, 20, 48, and $60~\text{keV}$. The RAR density profiles for M31 show the same three-region morphology as observed for the Milky Way \cite{2018PDU....21...82A}. In table \ref{table1}, we have the values for central densities ranging from $6.3\times10^{11}~M_{\odot}/\text{pc}^{3}$ at $6~\text{keV}$ to $1.6\times10^{20}~M_{\odot}/\text{pc}^{3}$ at $60~\text{keV}$. In figure \ref{fig:plot1}, we observe:
\begin{enumerate}
\item\textbf{Quantum Core}: a degenerate core with almost constant density and the degeneracy parameter $\theta(r)\gg1$, supported by the fermion degeneracy pressure and the particles are locked into a rigid quantum core-like structure.
\item\textbf{Transition Region}: outside the core, the density is observed dropping monotonically and the quantum effects start to fade. The degeneracy parameter transitions from positive to negative values, a transition from quantum core to classical halo.
\item\textbf{Classical Halo}: the classical regime, where quantum effects no longer matter and the particles follow Boltzmann statistics $\theta(r)<0$. It has a nearly constant density plateau, followed by a steep decline towards the boundary radius $r_{b}$ because of the cutoff.
\end{enumerate}

\subsection{Rotation Curves}
\begin{figure*}[t!]
\centering
\subfigure{\includegraphics[width=0.55\textwidth]{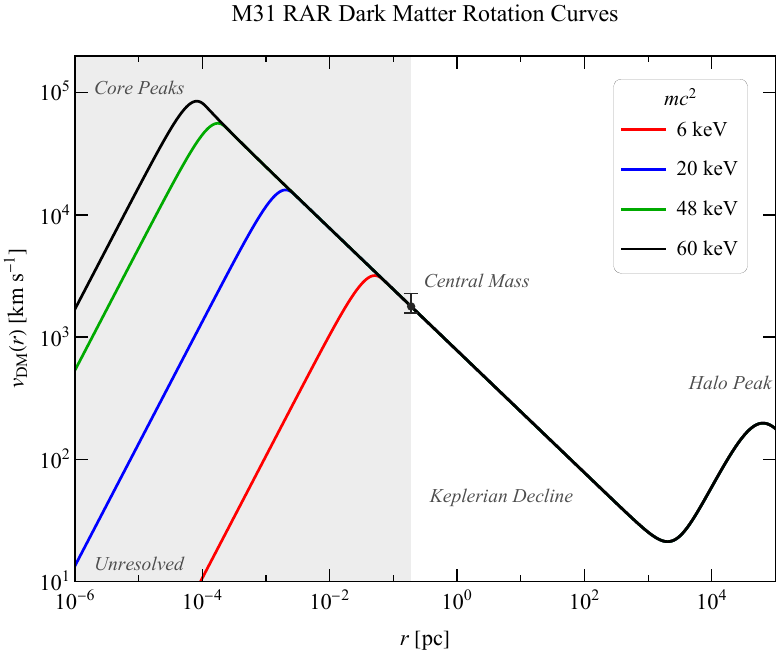}}
\caption{The M31 dark matter rotation curves for $mc^{2}=6~\text{keV}$ (solid red curve), $20~\text{keV}$ (solid blue curve), $48~\text{keV}$ (solid green curve), and $60~\text{keV}$ (solid black curve). Each curve shows a linearly increasing behavior till the first peak ($r_{c}$) i.e. the core radius, a Keplerian decline, rising again to the second peak ($r_{h}$) i.e. halo radius, and then decreasing toward the boundary set by the cutoff. The point at $0.19~\text{pc}$ shows the circular velocity because of the observed central mass, and the shaded region is marked for the radii below $0.19~\text{pc}$ which are not yet observationally resolved, hence the unresolved region.}
\label{fig:plot2}
\end{figure*}

To get the rotation curves, we plot the circular velocity as a function of radius $v_{\text{DM}}(r)$ given in Eq. \ref{eqn:DM}, for $mc^{2}=6$, 20, 48, and $60~\text{keV}$. The dark matter rotation curves for M31 show how the velocity varies from the center towards the edge of the galaxy. In figure \ref{fig:plot2}, we observe:
\begin{enumerate}
\item The velocity increases linearly, $v_{\text{DM}}\propto r$ up to the first maximum at $r_{c}$, the core radius \cite{2015MNRAS.451..622R}. 
\item Beyond the core, the velocity shows a Keplerian behavior, $v_{\text{DM}}\propto r^{-1/2}$ decreasing until it reaches a minimum $\approx 21~\text{km}~\text{s}^{-1}$ at $\approx2~\text{kpc}$, where the mass of the halo becomes comparable to that of the core. 
\item The velocity then rises to a second maximum of $\approx 197~\text{km}~\text{s}^{-1}$ at the halo radius $r_{h} \approx63.6~\text{kpc}$ \cite{2015MNRAS.451..622R}. Beyond $r_{h}$, it decreases, and the cutoff ends the halo at the boundary radius.
\end{enumerate}

The rotation curves for $mc^{2}=6\text{--}60~\text{keV}$ coincide beyond $\sim$ $0.1~\text{pc}$, while the core radius decreases as the fermion mass increases. Thus, the fermion mass sets the compactness of the core, while the extended dark matter halo remains almost unchanged because all the solutions satisfy the same boundary conditions. The resolved radius for M31's central mass is $0.19~\text{pc}$ \cite{2005ApJ...631..280B} and it is not further resolved for now, the shaded area below this radius is unresolved and the dark matter cores of all our solutions lie inside it.

\begin{table*}[t!]
	\centering
	\renewcommand{\arraystretch}{1.3}
	\setlength{\tabcolsep}{3pt}
	\begin{tabular}{lcccccc}
		\toprule[2pt]
		$m(\text{keV/c}^2)$ & $\beta_0$ & $\theta_0$ & $W_0$ & $\rho_{c}(M_{\odot}/\text{pc}^{3})$ & $r_{\text{c}}(\text{pc})$ & $v_{\text{c}}(\text{km}~\text{s}^{-1})$ \\
		\midrule
		6 & $4.39211\times 10^{-6}$ & 29.61092 & 52.39234 & $6.30444\times 10^{11}$ & $5.06991\times10^{-2}$ & 3186.41 \\
		10 & $1.66066\times 10^{-5}$ & 30.71614 & 54.17412 & $3.78012\times 10^{13}$ & $1.30317\times10^{-2}$ & 6308.67 \\ 
		20 & $1.02641\times 10^{-4} $& 32.13669 & 56.53963 & $9.98225\times 10^{15}$ & $2.04174\times10^{-3}$ & 16019.3 \\ 
		48 & $1.21511\times 10^{-3}$ & 34.63695 & 60.13303 & $1.58991\times 10^{19}$ & $1.75792\times10^{-4}$ & 56201.1 \\ 
		56 & $2.09731\times 10^{-3}$ & 35.66208 & 61.279400 & $7.28596\times 10^{19}$ & $1.05925\times 10^{-4}$ & 73800.0 \\ 
		60 & $2.81116\times 10^{-3}$ & 36.43144 & 62.07157 & $1.59484\times 10^{20}$  & $8.22243\times 10^{-5}$ & 85265.0
		\\
		\bottomrule[2pt]
	\end{tabular}
	\caption{Core properties of the RAR solutions for M31: free parameters of the model (the temperature parameter $\beta_{0}$, the degeneracy parameter $\theta_{0}$, and the cutoff parameter $W_{0}$), the central density $\rho_{c}$, the core size $r_{c}$, and the circular velocity $v_{c}$ at $r_{c}$, for different fermion masses.\label{table1}}
\end{table*}
The core velocities in table \ref{table1} ($3.19\times10^{3}\text{--}8.53\times10^{4}~\text{km}~\text{s}^{-1}$) are the dark matter circular velocities at the core radius $r_{c}\leq0.051~\text{pc}$. The observed circular velocity of M31 is approximately $1700~\text{km}~\text{s}^{-1}$ at the innermost observational limit on radius of $0.19~\text{pc}$ \cite{2005ApJ...631..280B}. All the solutions satisfy Eq. \ref{eqn:bc1}, and give circular velocity $v(0.19~\text{pc})\approx 1780~\text{km}~\text{s}^{-1}$ consistent with the observed value within the mass uncertainty ($1580\text{--}2280~\text{km}~\text{s}^{-1}$). At $31.23~\text{kpc}$, the dark matter circular velocity is $154~\text{km}~\text{s}^{-1}$ against the observed velocity of $204\pm38~\text{km}~\text{s}^{-1}$, at $98.74~\text{kpc}$, the dark matter velocity is $179~\text{km}~\text{s}^{-1}$, whereas the observed velocity is $193\pm42~\text{km}~\text{s}^{-1}$. The dark matter velocity stays below the observed values because the bulge and disk components are not included.

\subsection{Total Rotation Curve (Baryonic and Dark Matter)}
\begin{figure*}[t!]
\centering
\subfigure{\includegraphics[width=0.55\textwidth]{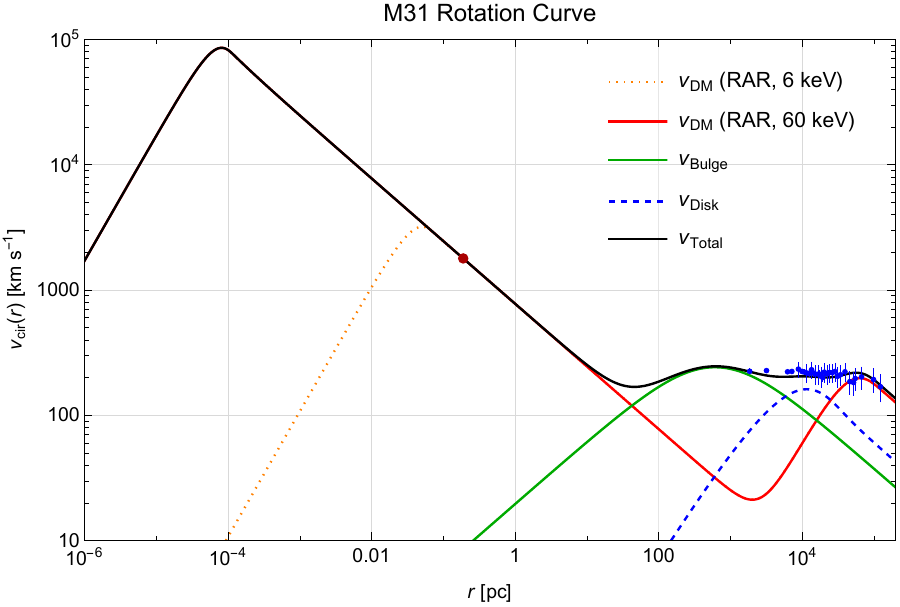}}
\caption{The M31 RAR model rotation curve, for $mc^{2}=6~\text{keV}$ (dotted orange curve) and $mc^{2}=60~\text{keV}$ (solid red curve), the bulge (solid green curve), the disk (dashed blue curve), and the total rotation curve (baryonic + dark matter) for $mc^{2}=60~\text{keV}$ (solid black curve). The bulge is a Hernquist profile and the disk a Freeman exponential profile. The model parameters for bulge and disk are from Geehan et al. \cite{2006MNRAS.366..996G}. The observational data points with error bars are from Zhang et al. \cite{2024MNRAS.528.2653Z}. The point with the error bar at $0.19~\text{pc}$ marks the central velocity given by the central mass $M=1.4^{+0.9}_{-0.3}\times10^{8}~M_{\odot}$ \cite{2005ApJ...631..280B},  and the shaded region shows the unresolved radii below $0.19~\text{pc}$.}\label{fig:plot3}
\end{figure*}

Baryonic matter also exists in galaxies; therefore, in figure \ref{fig:plot3}, we study the total rotation curve of M31 that includes baryonic and dark matter. To compare the model with the observed rotation curve, the observational data points of Zhang et al. \cite{2024MNRAS.528.2653Z} are added, and the best-fit parameters of Geehan et al. \cite{2006MNRAS.366..996G} are used for the baryons. The total circular velocity given in Eq. \ref{eqn:bardark} has three components: a Hernquist bulge \cite{1990ApJ...356..359H}, an exponential disk \cite{1970ApJ...160..811F}, and the RAR dark matter. The RAR dark matter comprises both the quantum core and the halo as a single continuous distribution. The velocity components are defined as

\begin{enumerate}
\item The bulge, which contains mostly old stars, is given by the Hernquist profile \cite{1990ApJ...356..359H}
\begin{equation}\label{eqn:bulge}
v_{\text{bulge}}^{2}=\frac{GM_{\text{b}}r}{(r+a_{\text{b}})^{2}},
\end{equation}
\item The exponential disk is given by Freeman \cite{1970ApJ...160..811F}
\begin{eqnarray}\label{eqn:disk}
v_{\text{disk}}^{2}&=&\frac{2GM_{\text{d}}}{a_{\text{d}}}y^{2}[I_{0}(y)K_{0}(y)-I_{1}(y)K_{1}(y)], \\
y&=&\frac{r}{2a_{\text{d}}},
\end{eqnarray}
where $I_{n}$ and $K_{n}$ are the modified Bessel functions of the first and second kind. We use the best-fit model of Geehan et al. \cite{2006MNRAS.366..996G} with the bulge mass $M_{\text{b}}=3.3\times10^{10}~M_{\odot}$ and scale radius $a_{\text{b}}=0.61\pm 0.10~\text{kpc}$, and the disk mass $M_{\text{d}}=8.4\times10^{10}~M_{\odot}$ and scale length $a_{\text{d}}=5.40\pm 0.13~\text{kpc}$.
\item The dark matter contribution, for core and halo, is given by the RAR dark matter circular velocity $v_{\text{DM}}$ of Eq. \eqref{eqn:DM} \cite{2015MNRAS.451..622R}. We have not modified any baryonic parameter in this work, apart from the test in section \ref{S4}. 
\end{enumerate}

\begin{figure*}[t!]
	\centering
	\subfigure{\includegraphics[width=0.55\textwidth]{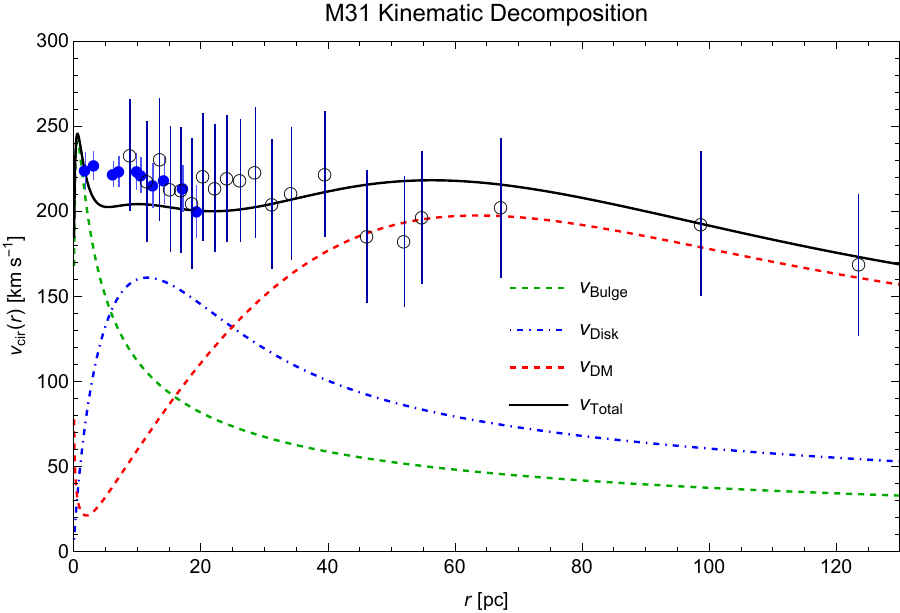}}
	\caption{The M31 total rotation curve on a linear scale for $mc^{2}=60~\text{keV}$ (solid black curve), with the bulge (dashed green curve), disk (dot dashed blue curve), and dark matter (dashed red curve). The observational data points along with their uncertainties added from Zhang et al. \cite{2024MNRAS.528.2653Z}, disk points (filled circles) and halo points (open circles).}\label{fig:plot4}
\end{figure*}	
The rotational curve data for M31 from Zhang et al. \cite{2024MNRAS.528.2653Z} has 20 disk points at $1.86\text{--}19.38~\text{kpc}$ with uncertainties of $7\text{--}15~\text{km}~\text{s}^{-1}$ and 20 halo points at $8.88\text{--}123.56~\text{kpc}$ with uncertainties of $32\text{--}42~\text{km}~\text{s}^{-1}$. The disk points come from the kinematics of disk stars and emission-line objects, whereas the halo points are obtained from Jeans modeling of halo stars, planetary nebulae, and star clusters \cite{2024MNRAS.528.2653Z}. We use all 20 halo points and 10 disk points at $r=1.86$, 3.19, 6.27, 7.20, 9.93, 10.62, 12.48, 14.29, 17.21, $19.38~\text{kpc}$. Figure \ref{fig:plot3} shows the dark matter rotation curves for $mc^{2}=6~\text{keV}$ and $mc^{2}=60~\text{keV}$ together with the bulge and disk profiles. It also shows the total rotation curve including the baryonic and dark matter, with the observational data points \cite{2024MNRAS.528.2653Z}. The bulge dominates the circular velocity within $5.9~\text{kpc}$, the disk dominates between $5.9$ and $25~\text{kpc}$, where it peaks at $161~\text{km}~\text{s}^{-1}$ at $11.6~\text{kpc}$, and the dark matter beyond  $25~\text{kpc}$. Beyond $28~\text{kpc}$, the dark matter exceeds the bulge and disk combined.

Figure \ref{fig:plot4} is the linear scale view of figure \ref{fig:plot3}. It shows the bulge, disk, dark matter, and total rotation curves with the disk points (filled circles) and the halo points (open circles). The model reproduces the halo points within their uncertainties ($\chi^{2}=5.0$ for 20 points) as well as the innermost and outermost disk points. The curve lies $\sim$$17\text{--}20~\text{km}~\text{s}^{-1}$ below the disk data at $\sim$$3\text{--}11~\text{kpc}$ and $\sim$$11\text{--}15~\text{km}~\text{s}^{-1}$ below at $\sim$$12\text{--}18~\text{kpc}$, where the baryons dominate. For instance, at $r=6.27~\text{kpc}$ the dark matter contributes $\sim$$39~\text{km}~\text{s}^{-1}$, whereas the bulge and disk contribute about 137 and $144~\text{km}~\text{s}^{-1}$, respectively, giving $v_{\text{tot}}=203~\text{km}~\text{s}^{-1}$, while the observed velocity is $222\pm7~\text{km}~\text{s}^{-1}$. At $3\text{--}11~\text{kpc}$, the dark matter provides only $11\text{--}31\%$ of the circular velocity, at these radii its profile is fixed by the boundary conditions at $31.23$ and $98.74~\text{kpc}$ and does not depend on the fermion mass. Since the baryons dominate these regions, this indicates that the velocity offset is primarily due to the adopted baryonic models rather than the dark matter profile.

\subsection{A Comparison Between the Milky Way and M31}
\begin{figure*}[t!]
	\centering
	\subfigure{\includegraphics[width=0.55\textwidth]{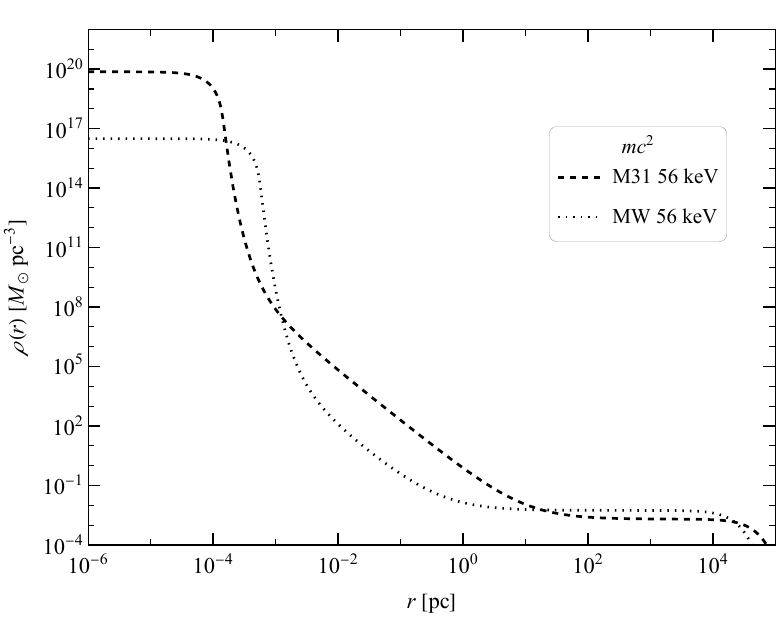}}
	\caption{The dark matter density profiles for M31 (black dashed curve) and the Milky Way (black dotted curve) for $mc^{2}=56~\text{keV}$. The parameters used for the Milky Way profile are from Becerra-Vergara et al. \cite{2020A&A...641A..34B}.}\label{fig:plot10}
\end{figure*}

\begin{figure*}[t!]
	\centering
	\subfigure{\includegraphics[width=0.55\textwidth]{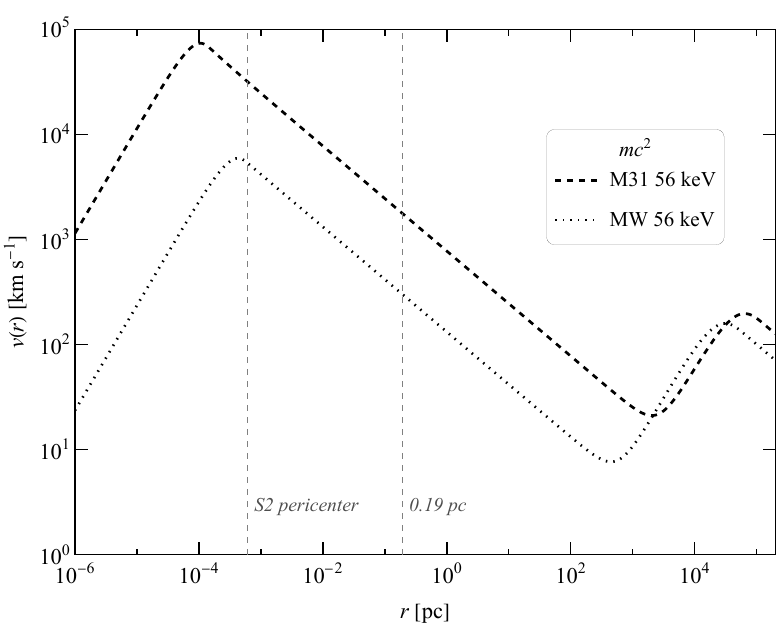}}
	\caption{The dark matter rotation curves for M31 (black dashed curve) and the Milky Way (black dotted curve) for $mc^{2}=56~\text{keV}$. The two vertical lines mark the S2 pericenter ($6\times10^{-4}~\text{pc}$) and the innermost resolved radius of M31 ($0.19~\text{pc}$). The parameters used for the Milky Way profile are from Becerra-Vergara et al. \cite{2020A&A...641A..34B}.}\label{fig:plot11}
\end{figure*}

We compare the profiles of the Milky Way and M31 for the same fermion mass, $mc^{2}=56~\text{keV}$, which is the best-fit value for the Milky Way \cite{2020A&A...641A..34B,Becerra-Vergara:2021gmx}. This value lies in the allowed range of both the galaxies, $mc^{2}=48\text{--}60~\text{keV}$. For the Milky Way, we use the parameters of Becerra-Vergara et al. \cite{2020A&A...641A..34B}, $\beta_{0}=1.1977\times10^{-5}$, $\theta_{0}=37.7656$, and $W_{0}=66.3407$, and for M31 we use the parameters from table \ref{table1}. In figure \ref{fig:plot10}, we see a comparison between the two density profiles which show that M31 has a more compact and dense core, and its core size is smaller than the size of the Milky Way's core. Figure \ref{fig:plot11} compares the rotation curves of both galaxies. To compare the two galaxies, we check their core and halo properties (given in table \ref{table3}): the central density, core mass, core radius, halo radius, core velocity, halo velocity, and core compactness.

\begin{table*}[t]
	\centering
	\renewcommand{\arraystretch}{1.3}
	\setlength{\tabcolsep}{8pt}
	\begin{tabular}{lcc}
		\toprule[2pt]
		& Milky Way & M31 \\
		\midrule
		$\rho_{c}(M_{\odot}/\text{pc}^{3})$ & $3.10\times10^{16}$ & $7.29\times10^{19}$  \\
		$M_{c}(M_{\odot})$ & $3.5\times10^{6}$ & $1.2\times10^{8}$  \\
		$r_{c}(\text{pc})$ & $4.28\times10^{-4}$ & $1.06\times10^{-4}$  \\
		$r_{h}(\text{kpc})$ & 30.9 & 63.6 \\
		$v_{c}(\text{km}~\text{s}^{-1})$ & 5937.2 & 73800.0 \\
		$v_{h}(\text{km}~\text{s}^{-1})$ & 158.85 & 197.45 \\
		$GM_{c}/(c^{2}r_{c})$ & 0.00039 & 0.054 \\
		\bottomrule[2pt]
	\end{tabular}
	\caption{Core and halo properties of the RAR solutions: central density $\rho_{c}$, core mass $M_{c}$, core radius $r_{c}$, halo radius $r_{h}$, core velocity $v_{c}$, halo velocity $v_{h}$, and core compactness $GM_{c}/(c^{2}r_{c})$ for the Milky Way and M31 for $mc^{2}=56~\text{keV}$.}\label{table3}
\end{table*}

The Milky Way and M31 both show the same core--halo morphology, the major difference lies in their cores. The core of M31 has a mass of about $1.2\times10^{8}~M_{\odot}$, which is 34 times that of the Milky Way core, $3.5\times10^{6}~M_{\odot}$; the remaining $2\times10^{7}~M_{\odot}$ of the central mass of M31 lies in the region between $r_{c}$ and $0.19~\text{pc}$. The central density of M31 is about $\rho_{c}=7.29\times10^{19}~M_{\odot}/\text{pc}^{3}$, which is roughly 2300 times that of the Milky Way, $\rho_{c}=3.10\times10^{16}~M_{\odot}/\text{pc}^{3}$. Although the core of M31 is more massive, its radius ($r_{c}=1.06\times10^{-4}~\text{pc}$) is about 4 times smaller than that of the Milky Way ($r_{c}=4.28\times10^{-4}~\text{pc}$). As for white dwarfs, a more massive degenerate core is smaller, because it must be compressed to a higher density for its degeneracy pressure to balance the strong gravity. The core compactness of M31, $GM_{c}/(c^{2}r_{c})=0.054$, is about 140 times that of the Milky Way, so the M31 core is much closer to gravitational collapse, which sets its upper bound to $mc^{2}\lesssim60~\text{keV}$.

The halos of the two galaxies are more alike than their cores. The two densities are nearly constant at about $1\text{--}10~\text{kpc}$, $1.9\times10^{-3}\text{--}2.1\times10^{-3}~M_{\odot}/\text{pc}^{3}$ for M31 and $4.1\times10^{-3}\text{--}5.5\times10^{-3}~M_{\odot}/\text{pc}^{3}$ for the Milky Way, before the cutoff takes effect. Figure \ref{fig:plot10} shows that the halo of M31 is more extended. Figure \ref{fig:plot11} shows the rotation curves: for M31, the curve peaks at $197~\text{km}~\text{s}^{-1}$ at $63.6~\text{kpc}$ and for the Milky Way, the curve peaks at $159~\text{km}~\text{s}^{-1}$ at $30.9~\text{kpc}$. The S2 star pericenter is about $6\times10^{-4}~\text{pc}$, the Milky Way core, with $r_{c}=4.3\times10^{-4}~\text{pc}$, lies inside it, which sets its lower bound, $mc^{2}\gtrsim48~\text{keV}$, whereas the core of M31, with $r_{c}=1.06\times10^{-4}~\text{pc}$, lies deep inside the resolved radius of $0.19~\text{pc}$. The mass ranges for the two galaxies overlap at $mc^{2}=48\text{--}60~\text{keV}$. The overlap is from the Milky Way range given by Argüelles et al. \cite{2018PDU....21...82A}. From the thermodynamic stability of Krut et al. \cite{Krut:2025jju}, we see a range of $mc^{2}\approx194\text{--}387~\text{keV}$ for the Milky Way. The critical mass at $194~\text{keV}$ is only about a tenth of its value at $60~\text{keV}$, where it equals the central mass of M31, and even lower at higher masses. No fermionic core can supply the central mass of M31 in this range, it would have to be a black hole, possibly formed  by collapse of a core exceeding the critical mass \cite{Arguelles:2023kqw}. For galaxies of about $10^{12}~M_{\odot}$ or more, Krut et al. \cite{Krut:2025jju} reached a similar conclusion. We have not yet tested the thermodynamic stability of our M31 solutions. For $mc^{2}=56~\text{keV}$, the stable solutions only exist for halos below about $10^{10}~M_{\odot}$ \cite{Krut:2025jju}, and the M31 halo has about $7.3\times10^{11}~M_{\odot}$ within $98.74~\text{kpc}$, so our solutions may not satisfy this additional condition.
	
\section{Statistical Kinematic Analysis}\label{S4}
We studied the theoretical framework of the RAR model for M31, now we need to test whether the model is consistent with the kinematics of M31 for the fermion masses $mc^{2}=6\text{--}60~\text{keV}$. We compare the total rotation curve with the observational data of Zhang et al. \cite{2024MNRAS.528.2653Z}. The data set has $N=30$ points of which $N_{\text{disk}}=10$ are the disk points ($1.86\text{--}19.38\ \text{kpc}$) and $N_{\text{halo}}=20$ are the halo points ($8.88\text{--}123.56\ \text{kpc}$). For each fermion mass, we compute the total $\chi^{2}$, the reduced $\chi^{2}/N_{\text{dof}}$ where $N_{\text{dof}}=N-k$ is the number of degrees of freedom and $k$ is the number of parameters fixed by these data, the $p$-value, and $\chi^{2}/N_{\text{disk}}$ and $\chi^{2}/N_{\text{halo}}$ for the disk and halo points. The total $\chi^{2}$ is defined as
\begin{equation}\label{eqn:chi}
\chi^{2}=\sum_{i=1}^{N}\left[\frac{v_{\text{obs}}(r_{i})-v_{\text{tot}}(r_{i})}{\sigma_{i}}\right]^2,
\end{equation}
where $v_{\text{obs}}(r_{i})$ is the observed circular velocity at radius $r_{i}$, $v_{\text{tot}}(r_{i})$ is the total modeled velocity, and $\sigma_{i}$ is the observational uncertainty. The baryonic parameters are fixed, and the three RAR parameters are set by the three boundary conditions. Two of the boundary conditions given in Eq. \ref{eqn:bc2}, come from the points of this data set, while the third, from the central mass in Eq. \ref{eqn:bc1}, is not a part of it. We therefore use $k=2$, i.e.\ $N_{\text{dof}}=N-2=28$ for the reduced $\chi^{2}$ and for the $p$-value, the probability of obtaining a $\chi^{2}$ larger than the observed one. The results are given in table \ref{table2}.
\begin{table*}[t]
	\centering
	\renewcommand{\arraystretch}{1.3}
	\setlength{\tabcolsep}{5pt}
	\begin{tabular}{lccccccc}
		\toprule[2pt]
		$m(\text{keV/c}^{2})$ & $M_{31.23}/M_{\text{bc}}$ & $M_{98.74}/M_{\text{bc}}$ & $\chi^{2}_{\text{tot}}$ & $\chi^{2}/N_{\text{dof}}$ & $p$-value & $\chi^{2}/N_{\text{disk}}$ & $\chi^{2}/N_{\text{halo}}$\\
		\midrule
		6  & 1.0000 & 1.0000 & 32.80 & 1.17 & 0.24 & 2.79 & 0.25 \\
		10 & 1.0000 & 1.0000 & 32.91 & 1.18 & 0.24 & 2.80 & 0.25 \\
		20 & 1.0000 & 1.0000 & 32.94 & 1.18 & 0.24 & 2.80 & 0.25 \\
		48 & 1.0000 & 0.9999 & 32.95 & 1.18 & 0.24 & 2.80 & 0.25 \\
		56 & 1.0000 & 1.0000 & 32.95 & 1.18 & 0.24 & 2.80 & 0.25 \\
		60 & 0.9998 & 0.9999 & 32.95 & 1.18 & 0.24 & 2.80 & 0.25 \\
		\bottomrule[2pt]
	\end{tabular}
	\caption{Statistical kinematic analysis of the RAR solutions for M31. $M_{31.23}$ and $M_{98.74}$ are the enclosed dark matter masses divided by their boundary values $M_{\text{bc}}$, Eq. \ref{eqn:bc2}. The total $\chi^{2}_{\text{tot}}$ is computed for $N=30$ points. The table also gives  the reduced $\chi^{2}/N_{\text{dof}}$, the $p$ value, and $\chi^{2}$ per data point for the disk and halo points.}\label{table2}
\end{table*}
The statistical analysis shows an agreement between the RAR framework and the observed kinematics of M31. For the different fermion masses, we have $\chi^{2}=32.80\text{--}32.95$, $\chi^{2}/N_{\text{dof}}=1.17\text{--}1.18$, and a $p$-value of $0.24$. Since all the solutions share the same halo, $\chi^{2}$ changes by only $0.15$ from $6\text{--}60~\text{keV}$, so the rotation curve of M31 does not single out a fermion mass. For the halo, $\chi^{2}/N_{\text{halo}}\approx0.25$ which means $\chi^{2}=5.0$ for the 20 halo data points, of which the two used as boundary conditions are matched exactly, the model lies well within the uncertainties of the halo points. The model lies $17\text{--}20~\text{km}~\text{s}^{-1}$ below the data points at $3\text{--}11~\text{kpc}$, this offset results in the disk $\chi^{2}/N_{\text{disk}}\approx2.8$, that is, $\chi^{2}=28$ for 10 disk data points. In this region, the dark matter contributes only about $11\text{--}31\%$ of the circular velocity. A disk heavier by $20\%$ would raise $v_{\text{disk}}$ by 10\% (less than the change of up to 15\% in the disk velocity adopted for the Milky Way in Ref. \cite{2018PDU....21...82A}), this reduces the offset to $6\text{--}14~\text{km}~\text{s}^{-1}$ and the disk $\chi^{2}$ from 28 to 7.9, with the boundary conditions re-derived for this disk. Using all 20 disk points, we have $\chi^{2}/N_{\text{dof}}\approx 1.50$ ($p=0.24$), and $p=0.999$ with a disk 20\% more massive. Therefore, the offset is attributed to the adopted baryonic profiles and their parameters rather than to the fermion dark matter distribution.

\section{Comparison With the Navarro--Frenk--White and Burkert Profiles}\label{S5}
We compare the RAR profiles with the Navarro--Frenk--White (NFW) \cite{Navarro:1995iw} (cuspy) and Burkert \cite{Burkert:1995yz} (cored) profiles to get a better understanding of the results. The NFW profile is the most widely used dark matter profile. The NFW and Burkert profiles are
\begin{equation}\label{eqn:nfw}
\rho_{\text{NFW}}(r)=\frac{\rho_{s}}{(\frac{r}{r_{s}})(1+\frac{r}{r_{s}})^{2}},
\end{equation}
\begin{equation}\label{eqn:bur}
\rho_{\text{Bur}}(r)=\frac{\rho_{0} r_{0}^{3}}{(r+r_{0})(r^{2}+r_{0}^{2})},
\end{equation}
where $r_{s}$ is the scale radius of the NFW profile, at which the logarithmic slope of the density is $-2$, $\rho_{s}$ is the characteristic density, $\rho_{0}$ is the central density, and $r_{0}$ is the core radius. These profiles describe the dark matter halos, not accounting for the central mass, we therefore add a black hole of $1.4\times10^{8}~M_{\odot}$ to both the profiles, unlike the RAR model in which the fermions that provide the halos also provide the central core. We use the same Hernquist bulge and Freeman exponential disk as we used for the RAR model, with the parameters of Geehan et al. \cite{2006MNRAS.366..996G}. The NFW and Burkert circular velocities are
\begin{equation}\label{eqn:nfwv}
v_{\text{NFW}}^{2}(r)=\frac{G(M_{\text{NFW}}+M_{\text{BH}})}{r},
\end{equation}
\begin{equation}\label{eqn:burv}
v_{\text{Bur}}^{2}(r)=\frac{G(M_{\text{Bur}}+M_{\text{BH}})}{r},
\end{equation}
where $M_{\text{NFW}}$ and $M_{\text{Bur}}$ are the enclosed dark matter masses for NFW and Burkert profiles and $M_{\text{BH}}$ is the black hole mass. The total rotation curves will be
\begin{equation}\label{eqn:tot2}
v_{\text{tot}}^{2}(r)=v_{\text{bulge}}^{2}(r)+v_{\text{disk}}^{2}(r)+v_{\text{NFW}}^{2}(r),
\end{equation}
\begin{equation}\label{eqn:tot3}
v_{\text{tot}}^{2}(r)=v_{\text{bulge}}^{2}(r)+v_{\text{disk}}^{2}(r)+v_{\text{Bur}}^{2}(r).
\end{equation}
We fit both the profiles to the 30 data points of Zhang et al. \cite{2024MNRAS.528.2653Z}, 10 disk points and 20 halo points, by minimizing $\chi^{2}$, Eq. \ref{eqn:chi}. The best-fits give $r_{s}=42.1~\text{kpc}$, $\rho_{s}=1.36\times10^{-3}~M_{\odot}~\text{pc}^{-3}$ for the NFW profile, with $\chi^{2}=5.56$ and $r_{0}=13.1~\text{kpc}$ and $\rho_{0}=1.25\times10^{-2}~M_{\odot}~\text{pc}^{-3}$ for the Burkert profile, with $\chi^{2}=11.82$. Both fit the data better than the RAR solutions, where the RAR halo is fixed by the boundary conditions rather than fitted to the rotation curve like the NFW and Burkert profiles (the difference mainly lies in the disk points).
\begin{figure*}[t!]
	\centering
	\subfigure{\includegraphics[width=0.55\textwidth]{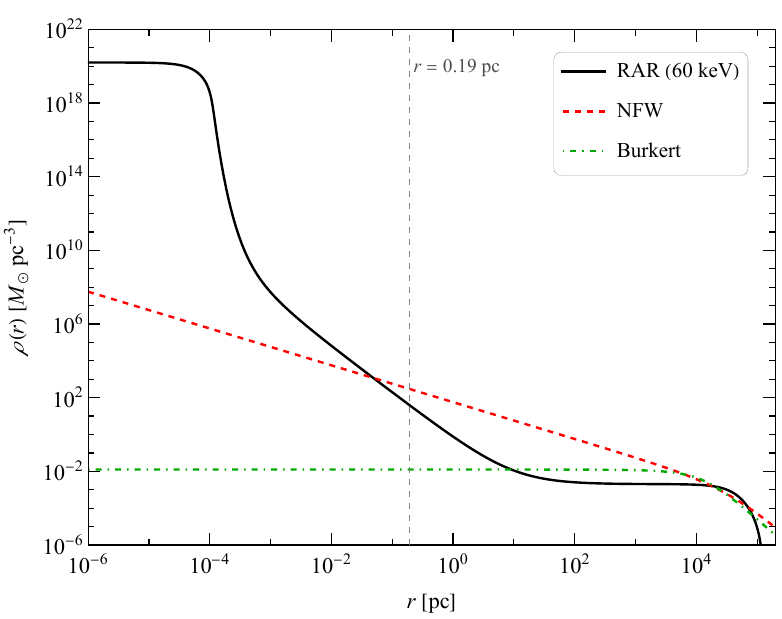}}
	\caption{The M31 dark matter density profile for $mc^{2}=60~\text{keV}$ (solid black curve), the fitted NFW (red dashed curve), and Burkert (green dot dashed curve) densities are plotted to get a comparison among the three models. The dotted vertical line shows the observational point at $0.19~\text{pc}$, the innermost radius at which the central mass is measured \cite{2005ApJ...631..280B}.}\label{fig:plot8}
\end{figure*}
\begin{figure*}[t!]
	\centering
	\subfigure{\includegraphics[width=0.55\textwidth]{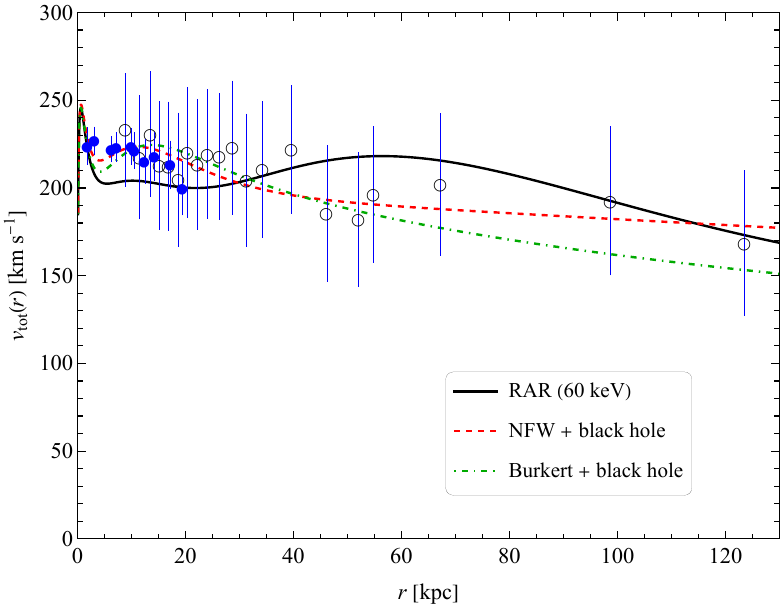}}
	\caption{The M31 total rotation curve for the RAR model $mc^{2}=60~\text{keV}$ (solid black curve), the fitted NFW (red dashed curve) and Burkert (green dot dashed curve) rotation curves are plotted to get a comparison among the three models. The bulge (Hernquist) and the disk (Freeman) are the same \cite{2006MNRAS.366..996G}, observational points along with uncertainties are added from Zhang et al. \cite{2024MNRAS.528.2653Z}. The NFW and Burkert profiles are halo models, so we add a black hole to them.}\label{fig:plot9}
\end{figure*}
Figure \ref{fig:plot8} shows the RAR dark matter density for $mc^{2}=60~\text{keV}$ along with the NFW and Burkert densities. The NFW density profile shows a spike going to infinity towards the center ($\rho \propto r^{-1}$). This cusp, which is not seen in many dwarf galaxies but is predicted by CDM simulations, is the famous core-cusp problem \cite{2017ARA&A..55..343B}; however, at large radii, the profile describes the halos well. The Burkert density stays nearly constant inside the core radius $r_{0}$, $\rho \to \rho_{0}$ as $r\to 0$. Both the profiles fall as $\rho \propto r^{-3}$ at large radii. The RAR profile gives $1.4\times10^{8}~M_{\odot}$ within $0.19~\text{pc}$, whereas the NFW and Burkert profiles give  masses of 13 and $4\times10^{-4}~M_{\odot}$, respectively. Figure \ref{fig:plot9} shows the total rotation curves of the RAR, NFW, and Burkert models. They differ mainly in the region $3\text{--}11~\text{kpc}$, where the NFW and Burkert profiles contain more dark matter than the RAR profile. At $8.2~\text{kpc}$, the dark matter velocities for the NFW and Burkert profiles are about $100~\text{km}~\text{s}^{-1}$ and  $92~\text{km}~\text{s}^{-1}$, respectively, whereas the RAR gives $49~\text{km}~\text{s}^{-1}$. The offset in the RAR curve can therefore be removed either with more dark matter, as seen in the NFW and Burkert fits, or with a heavier disk. 

Other dark matter models consider weakly interacting massive particles (WIMPs) or axions, self-interacting particles, and ultralight bosons as dark matter candidates. The CDM model considers particles such as WIMPs or axions which have not yet been detected \cite{Schumann:2019eaa,Irastorza:2018dyq}, and explains the large-scale structure of the Universe well, but on galactic scales it faces problems such as the core-cusp problem \cite{2017ARA&A..55..343B}. Warm dark matter consists of particles with masses of a few keV, such as sterile neutrinos \cite{Boyarsky:2018tvu}, RAR fermions in a certain keV range could also be sterile neutrinos \cite{2016JCAP...04..038A}. These particles stream freely in the early Universe, erasing the small-scale density fluctuations and reducing the number of small halos and satellite galaxies \cite{Boyarsky:2018tvu}. In self-interacting dark matter, the particles collide and scatter off each other, turning the central cusp into a core \cite{Spergel:1999mh,Tulin:2017ara}. Ultralight bosons with masses of about $10^{-22}~\text{eV}$ form fuzzy dark matter, these bosons may form a solitonic core at the center because of their quantum wave nature on galactic scales \cite{Hu:2000ke,Hui:2016ltb,Schive:2014dra}, whereas the self-interacting bosons form a Bose--Einstein condensate with a cored halo \cite{Boehmer:2007um}. These models differ mainly at the centers of galaxies.

\section{Summary and Discussion}\label{S6}
In this work, we applied the RAR model to M31 for fermion masses in the range $mc^{2}=6\text{--}60~\text{keV}$. The upper mass limit of this range comes from the core stability against the gravitational collapse, for $mc^{2}\lesssim60~\text{keV}$, the central mass is below the critical mass for the gravitational collapse. On solving the equilibrium equations of the RAR model numerically, we see a three-region morphology: a central degenerate quantum core, a transition region, and an external classical halo. At $6~\text{keV}$, we obtain a core radius of $0.051~\text{pc}$, which decreases to $8.2\times10^{-5}~\text{pc}$ at $60~\text{keV}$, all the cores lie within $0.19~\text{pc}$, the innermost observed radius. 

The solutions are consistent with the observed central kinematics, giving a circular velocity of $v\simeq1780~\text{km}~\text{s}^{-1}$ at $0.19~\text{pc}$, compared with the observed velocity of $v\simeq1700~\text{km}~\text{s}^{-1}$. We obtain core velocities from $3.2\times10^{3}~\text{km}~\text{s}^{-1}$ to $8.5\times10^{4}~\text{km}~\text{s}^{-1}$, these are the predicted values that current observations cannot yet test. By incorporating unadjusted baryonic components (a Hernquist bulge and a Freeman disk) \cite{2006MNRAS.366..996G}, we obtain a total rotation curve in agreement with the M31 data, with $\chi^{2}/N_{\text{dof}}=1.17\text{--}1.18$ and a $p$-value of 0.24 (section \ref{S4}). The curve reproduces the halo points within the observational uncertainties, however, we see an offset of $17\text{--}20~\text{km}~\text{s}^{-1}$ at $3\text{--}11~\text{kpc}$ below the disk points. This offset is attributed to the baryonic model of Geehan et al. \cite{2006MNRAS.366..996G} used in this work because the baryons dominate at these radii. A disk 20\% more massive reduces this offset (section \ref{S4}), but we prefer the unadjusted model and parameters.   

The solutions contain the central mass of $\sim1.4\times10^8~M_{\odot}$ within the innermost observational radius of $0.19~\text{pc}$. Furthermore, the fermion mass ranges for M31 and the Milky Way overlap at $mc^{2}=48\text{--}60~\text{keV}$ with the earlier Milky Way range \cite{2018PDU....21...82A}, which supports the universality of the RAR model that a single dark matter particle can describe the dark matter distribution in both galaxies. The critical mass falls below the central mass of M31 with the thermodynamic range of Krut et al. \cite{Krut:2025jju}, $mc^{2}\approx 194\text{--}387~\text{keV}$, so no fermionic core can supply it, and M31 would have to host a black hole. We compare the solutions for $mc^{2}=56~\text{keV}$ and find that the cores of the two galaxies differ greatly but the halos are more alike. We find that the M31 core is more compact and is 34 times more massive than the Milky Way core but 4 times smaller in size, hence the M31 core is much closer to gravitational collapse. The application of this framework to M31 provides a unified description of the two major astrophysical problems: it describes the dark matter distribution consistent with the central mass concentration, offering a channel for the formation of supermassive black holes, while simultaneously explaining the extended rotation curve of M31. We also compared the density profile and total rotation curve of the RAR model with the NFW and Burkert density profiles and rotation curves, using the same bulge and disk models and parameters \cite{2006MNRAS.366..996G}. We fitted the same data of Zhang et al. \cite{2024MNRAS.528.2653Z} to the NFW and Burkert rotation curves, these profiles give better fits than the RAR model, whose halo is fixed by the boundary conditions rather than fitted to the rotation curve. The NFW and Burkert profiles have masses of 13 and $4\times10^{-4}~M_{\odot}$ within $0.19~\text{pc}$ and hence require a separate black hole.

This innermost constraint at $0.19~\text{pc}$ may be refined in the future by high-resolution kinematic observations of M31's core. A few directions for future work include testing the thermodynamic stability of our M31 solutions, performing stability analyses under baryonic accretion to investigate core collapse and supermassive black hole formation, extending the model to other galaxies with well-measured central masses and high-quality rotation curve data, and evaluating the outer dark matter halo beyond the current boundary.

\bibliography{refs.bib}
\bibliographystyle{apsrev4-2}
	
\end{document}